\documentclass[a4paper,12pt]{article}
\usepackage{amsmath}
\usepackage{amssymb}
\usepackage{revsymb}
\usepackage{mathrsfs}
\begin{document}
\title{Pair production in a strong time-depending magnetic field: the effect of a strong gravitational field}
\author{Antonino Di Piazza\footnote{E-mail: dipiazza@mpi-hd.mpg.de}\\
Max-Planck-Institut f\"ur Kernphysik, \\
Saupfercheckweg 1, Heidelberg, D-69117, Germany
\\
\\
\and Giorgio Calucci\footnote{E-mail: giorgio@ts.infn.it}\\ 
Dipartimento di Fisica Teorica\\ 
Strada Costiera 11, Trieste, I-34014, Italy\\ 
and\\
INFN, Sezione di Trieste, Italy}
\maketitle
\begin{abstract}
We present the calculation of the probability production of an 
electron-positron pair in the presence of a strong magnetic 
field with time-varying strength. The calculation takes into 
account the presence of a strong, constant and uniform gravitational 
field in the same direction of the magnetic field. The results show 
that the presence of the gravitational field in general enhances very much the 
production of pairs. In particular, high-energy pairs are more likely produced 
in the presence of the gravitational field than in Minkowski spacetime.

\end{abstract}
\clearpage
\section{Introduction}
We have already calculated the production probability of 
$e^-\text{-}e^+$ pairs in the presence of an overcritical
($\gg B_{cr}=m^2c^3/\hbar e
\simeq 4.4\cdot 10^{13}\; \text{G}$ with $m$ and $-e<0$ the mass and the electric charge of the electron respectively), uniform and time varying 
magnetic field \cite{Calucci,DiPiazza1,DiPiazza2} by studying different kinds of 
time dependence such as the rotating magnetic field 
or the fixed-direction magnetic field with time-depending strength. 
Magnetic fields with these characteristics may be present 
around highly magnetized neutron stars (\textit{magnetars}) or black holes 
\cite{Thompson,Hanami,Kouveliotou} (obviously, the spatial uniformity of the 
magnetic field is assumed to hold in microscopic 
length scales of the order of the Compton length) where, on the other hand, 
it is generally believed that gamma-ray bursts are produced [the literature on gamma-ray bursts is endless, we quote the recent reviews \cite{GRB}]. 
Concerning this fact, the production of electrons 
and positrons is in our mind an intermediate step toward the production of photons 
through $e^-\text{-}e^+$ annihilations \cite{DiPiazza3} or bremsstrahlung \cite{DiPiazza5}, 
because the photons give the only experimental signal that can be observed. It is also worth stressing that our approach is always ``microscopic'' that means that we do not consider macroscopic effects like the detailed form of the magnetic field around the astrophysical compact object or the presence of other particles or of plasmas in the production region.

Now, it is clear that in the astrophysical scenario we refer to the 
gravitational field produced by the compact object 
(neutron star or black hole) can play an important role. As general aim 
of our investigation we consider the production of particles by a non 
stationary magnetic field, so we are interested in situations where the 
gravitational effects are not the dominant dynamical feature. However, even with 
these limitations we find interesting to consider situations where the 
gravitational field is strong enough to put into the game some aspects of 
general relativity. Consequently, the simplest realistic configuration we may find, i.e. 
the Schwarzschild metric, was chosen as starting point and then, analogously 
to the magnetic field, the gravitational field was taken as uniform 
over a Compton wavelength of the produced particle. If one is not too close to 
the event horizon of the compact object 
the gravitational effects may be treated perturbatively and 
this was already done in \cite{DiPiazza4}. In order to complete the investigation we suppose here that 
the particle production takes place near 
the event horizon where, because of the singularity of the spacetime 
metric, a perturbative approach is inapplicable. By restricting 
our attention to the pair production around black holes, 
the investigation is still possible because we can approximate 
the Schwarzschild metric in a form [the Rindler metric \cite{Rindler}], where 
the general covariant Dirac equation is solvable, 
even in the presence of a uniform magnetic field, provided 
the magnetic and the gravitational fields are parallel \cite{Torres,Bautista}. In the previous paper \cite{DiPiazza4} where we have treated perturbatively the gravitational field we had the possibility to choose more freely the mutual configuration between the two fields. The calculations are performed by using the quantum 
field theory in curved spacetimes \cite{Birrell,Fulling,Wald} in order to include 
the presence of the background gravitational field and the adiabatic 
perturbation theory \cite{Migdal} to deal with the time dependent magnetic 
field. In fact, the presence of the gravitational 
field is taken into account only in the calculation of the one-particle 
electron and positron states without considering the particle production induced 
by the gravitational field itself as in \cite{Hawking1,Hawking2,Stellare,Cosmologico} where the authors consider the production of particles by a time depending stellar \cite{Hawking1,Hawking2,Stellare} or cosmological \cite{Cosmologico} gravitational field.

The paper is structured as follows. In the next Section, starting from the Schwarzschild metric we expand it around the event horizon getting in this way a Rindler metric where a particular magnetic field is introduced. In Section \ref{III} the Hamiltonian, the one-particle eigenstates of a Dirac particle are displayed, with a particular attention to the energy spectrum which has, as expected, very different features with respect to the case where only the magnetic field is present. Finally, in Section \ref{IV} the pair production due to the time dependence of the magnetic field (in the constant gravitational field) is calculated and some conclusions are presented in Section \ref{V}. An appendix 
contains mathematical details of some results only quoted in the main text. To make easier the reading of the paper some intermediate calculations have been omitted: more details can be found in ArXiv under the number hep-ph/0406251.

Natural units ($\hbar=c=1$) are used below and the Minkowski spacetime metric 
is $\eta_{\mu\nu}=\text{diag}(+1,-1,-1,-1)$ with $\mu,\nu=0,1,2,3$ (Greek indices 
are supposed to run from $0$ to $3$ while Latin indices from $1$ to $3$).
%
%
\section{Theoretical model}
\setcounter{equation}{0}
\renewcommand{\theequation}{II.\arabic{equation}}
This Section is devoted to the description of the theoretical model within which 
our calculations are performed. As we have said in the Introduction, both the gravitational and 
the magnetic fields are assumed to be classical (not quantized) fields 
with a given temporal evolution. On the contrary, in order 
to describe the pair creation process, the electron-positron 
Dirac field has to be quantized. Even if we assume that the spacetime structure of the gravitational and the magnetic fields is given, it is very difficult to determine them because the system
built up by the Einstein equations and the general covariant Maxwell
equations should be solved. To the sake of clarity, we remind that we want to calculate the production probability of an electron-positron pair in the presence of a static strong gravitational field and of a time-varying strong magnetic field. In principle, a possible physical system that can give rise to these kinds of fields is represented by a rotating black-hole surrounded by a magnetized accretion disk\footnote{Because of the so called ``no-hair'' theorem [see e. g. \cite{Heusler}], an isolated black hole cannot generate a time-depending magnetic field because its behaviour is completely determined only by its mass, charge and angular momentum. In the most general case, a charged rotating black hole can generate a \emph{static} dipole magnetic field.}. In fact, this system is believed to be one of the possible candidates as a central engine of the gamma-ray bursts and the presence of a strong ($\sim 10^{15}\;\text{G}$) magnetic field makes possible the energy ``extraction'' from the black hole through the Blandord-Znajek mechanism \cite{Blandford,Lee_GRB,Barbiellini,Usov}. In particular, the capture of a near star by the rotating black hole can lead to the creation of the accretion disc and to the production of a strong transient magnetic field \cite{Cheng} [see also the Ref. \cite{Li} where the time evolution of the magnetic field around a rotating black hole is explicitly discussed]. Of course, it is impossible to determine analytically the gravitational and the electromagnetic field of this physical system. For this reason we will make a number of simplifying assumptions to proceed. First, since we imagine the pair to be produced near the black hole event horizon we can neglect the gravitational field produced
by the accretion disk and by the magnetic field and then we can assume that the spacetime metric is determined only by the black hole. Even if the black hole is rotating we can limit ourselves to study the production of pairs along (or near) the black hole rotational axis in order to neglect the effects of the rotation on the spacetime metric. In this simplified scenario our starting point is, as in \cite{DiPiazza4}, the metric tensor of a non-rotating spherical body of mass $M$ outside the body itself. If we indicate with $t$, $X$, $Y$ and $Z$ the so-called isotropic coordinates, this metric tensor can be written as \cite{Landau2}
\begin{equation}
\label{g_mu_nu}
g_{\mu\nu}(X,Y,Z)=\text{diag}\left[\frac{F^2_-(X,Y,Z)}
{F^2_+(X,Y,Z)},-F^4_+(X,Y,Z),-F^4_+(X,Y,Z),-F^4_+(X,Y,Z)\right]
\end{equation}
where
\begin{equation}
\label{F_pm}
F_{\pm}(X,Y,Z)=1\pm\frac{r_G}{4\sqrt{X^2+Y^2+Z^2}}
\end{equation}
with $r_G=2GM$ the gravitational radius of the body and $G$ the gravitational constant. We have chosen the isotropic metric instead of the usual and equivalent Schwarzschild metric because from Eq. (\ref{g_mu_nu}) we see
that the spatial metric is proportional to the Euclidean one and this will simplify our future calculations. In particular, if we imagine to pass to spherical coordinates \cite{Landau2} and we consider two points $P_j=(X_j,Y_j,Z_j)$ with $j=1,2$ that have the same angular coordinates but such that $R_j=\sqrt{X_j^2+Y_j^2+Z_j^2}$ and $R_2>R_1$ then the physical distance $\Delta l$ between $P_1$ and $P_2$ is given by
\begin{equation}
\Delta l=\int_{R_1}^{R_2}dR\left(1+\frac{r_G}{4R}\right)^2=\Delta R+\frac{r_G}{2}\log\left(\frac{R_2}{R_1}\right)+\left(\frac{r_G}{4}\right)^2\frac{\Delta R}{R_1R_2}
\end{equation}
with $\Delta R=R_2-R_1$.

Now, the $e^-\text{-}e^+$ pair production is a microscopic process which takes place in a volume 
with typical linear length of the order of the Compton length $\lambdabar=1/m$. In this length-scale the gravitational field 
produced by a macroscopic object does not 
vary very much and the form of the metric tensor (\ref{g_mu_nu}) can be simplified. 
As we have said in the Introduction, we 
want to consider here the case in which the pair is created  near the black hole event horizon lying at 
$\sqrt{X^2+Y^2+Z^2}=r_G/4$ \footnote{We do not consider the 
particular case in which the pair is created just on the event horizon 
of the black hole because in this case 
the direct particle production induced 
by the gravitational field can be the dominating 
production process and here we are not interested in it. Since the process of the direct production of a \emph{massive} particle can be interpreted as a tunnel effect it is sensible to affirm that it is relevant at distances from the event horizon of the order of the Compton wavelength of the created particle.}. We choose the reference system in 
such a way the pair is created in a volume centered on the $z$-axis then 
the previous considerations allow us to expand the metric tensor 
(\ref{g_mu_nu}) around the point $P_0=(0,0,r_G/4)$. If $P=(x,y,r_G/4+z)$ with 
$z>0$ is a generic point near $P_0$ then
\begin{align}
g_{00}(P) &=\left(\frac{2z}{r_G}\right)^2+O\left[\left(\frac{z}{r_G}\right)^3\right], && \\
g_{ii}(P) &=-16+O\left(\frac{z}{r_G}\right) && i=1,2,3.
\end{align}
It is clear that we are only interested in the pairs created in the $(z>0)$-region 
because those created in the $(z<0)$-region will fall into the black hole. Concerning the footnote 2 it is also worth stressing that in order that our treatment holds the quantity $z$ is not restricted to be microscopic: it is important only that $z\ll r_G$.

Now, if we keep only the lowest order non-zero term in $g_{\mu\mu}(P)$ then the initial 
metric tensor (\ref{g_mu_nu}) can be written approximately as
\begin{equation}
\label{g_R}
g_{\mu\nu}(P)\simeq g_{\mu\nu}^{(R)}(z)=\text{diag}
\left[\left(\frac{2z}{r_G}\right)^2,-16,-16,-16\right].
\end{equation}
This metric tensor has the same form of a Rindler metric tensor describing 
an observer uniformly accelerated in the $z$ direction \cite{Rindler}
\footnote{We could have scaled the spatial coordinates in 
order to have exactly a Rindler metric tensor, but we prefer to 
work with $x$, $y$ and $z$ that are the Cartesian coordinates at infinity.}. Actually, the physical meaning of our coordinates is 
very different from that of the coordinates in the Rindler spacetime. For example, while here the coordinate $t$ is precisely the time coordinate in the region far from the black hole, the time coordinate in the Rindler spacetime is a combination of the Minkowski time coordinate and of the Minkowski spatial coordinate along the acceleration. Nevertheless, the fact that the two metric tensors have the same form allows us to conclude that the metric tensor (\ref{g_R}) describes a 
constant and uniform gravitational field in the $z$ direction. Observe that 
no assumption is needed about the strength of the gravitational field itself.

Now, we pass to the mathematical description of the magnetic field. We will deal with a 
uniform magnetic field $\mathbf{B}(t)$ with constant direction and time-varying 
strength:
\begin{equation}
\label{B}
\mathbf{B}(t)=
\left(\begin{array}{c}
0\\
0\\
B(t)
\end{array}\right).
\end{equation}
As we have said in the Introduction, we 
consider this particular case of a magnetic field in the same direction 
of the gravitational field because only in this case the general covariant 
Dirac equation has been solved explicitly \cite{Torres,Bautista} and, in order to apply the adiabatic perturbation theory, we need the exact solution of that equation \cite{Migdal}.  Also, the magnetic field is considered to be strong ($B(t)\gg B_{cr}=m^2/e$) and 
slowly varying ($\dot{B}(t)/B(t)\ll \lambdabar^{-1}=m$). We choose the electromagnetic gauge in which 
the vector potential $A_{\mu}(\mathbf{r},t)$ that gives rise to $\mathbf{B}(t)$ 
is given by
\begin{align}
A_0(\mathbf{r},t) &=0,\\
A_1(\mathbf{r},t) &=\frac{1}{2}\left[\mathbf{r}\times\mathbf{B}(t)\right]_x=
\frac{yB(t)}{2},\\
A_2(\mathbf{r},t) &=\frac{1}{2}\left[\mathbf{r}\times\mathbf{B}(t)\right]_y=
-\frac{xB(t)}{2},\\
A_3(\mathbf{r},t) &=\frac{1}{2}\left[\mathbf{r}\times\mathbf{B}(t)\right]_z=0.
\end{align}
In the following we will also use the three-dimensional vector 
$\mathbf{A}(x,y,t)=(A_x(y,t),A_y(x,t),0)$ with
\begin{align}
\label{A_1}
A_x(y,t) &=-\frac{yB(t)}{2},\\
\label{A_2}
A_y(x,t) &=\frac{xB(t)}{2}
\end{align}
which is such that $\boldsymbol{\partial}\times \mathbf{A}(x,y,t)=\mathbf{B}(t)$. 

In order to calculate the pair production probability, 
we have to build the second quantized Hamiltonian 
of a Dirac field $\Psi (\mathbf{r},t)$ in the presence of the already 
introduced gravitational and magnetic 
fields. We start by writing the Lagrangian density of this system which is given 
by \cite{Birrell}
\begin{equation}
\label{L}
\begin{split}
\mathscr{L}^{(R)}(\Psi,\partial_{\mu}\Psi,\bar{\Psi},\partial_{\mu}\bar{\Psi},\mathbf{r},t)=
\sqrt{-g^{(R)}(x)}\bigg\{
&\frac{1}{2}\Big[\bar{\Psi}\gamma^{(R)\,\mu}(z)[i\partial_{\mu}+i\Gamma^{(R)}_{\mu}(z)+
eA_{\mu}(\mathbf{r},t)]\Psi-\\
&\left.-\bar{\Psi}[i\overleftarrow{\partial}_{\mu}-i\Gamma^{(R)}_{\mu}(z)-
eA_{\mu}(\mathbf{r},t)]\gamma^{(R)\,\mu}(z)\Psi\right]-\\
&-m\bar{\Psi}\Psi\bigg\}
\end{split}
\end{equation}
where $g^{(R)}(z)\equiv\det[g^{(R)}_{\mu\nu}(z)]$ is the determinant of the metric tensor, $\gamma^{(R)\,\mu}(z)\equiv\gamma^{\alpha}e_{\alpha}^{(R)\mu}(z)$ are the covariant Dirac matrices with $e_{\alpha}^{(R)\mu}(z)$ the tetrad field \cite{Birrell,Weinberg} and $\Gamma^{(R)}_{\mu}(z)\equiv-i\sigma^{\alpha\beta}e_{\alpha}^{(R)\nu}(z)
e_{\beta\nu;\mu}^{(R)}(z)/4$ are the so-called spin connections. The tetrad field $e_{\alpha}^{(R)\mu}(z)$ has been 
chosen to be diagonal with
\begin{align}
\label{V_0^(1)1}
e_0^{(R)0}(z) &=\frac{r_G}{2z},\\
e_i^{(R)i}(z) &=\frac{1}{4} && \text{no sum}.
\end{align}
With this choice the spatial spin connections $\Gamma^{(R)}_i(z)$ vanish 
while $\Gamma^{(R)}_0(z)$ is independent of $z$ and is given by $\Gamma^{(R)}_{0}=\gamma^0\gamma^3/4r_G$.

The Hamiltonian density is defined as \cite{Fulling_p_2}
\begin{equation}
\label{H_def}
\mathscr{H}^{(R)}(\Psi,\partial_i\Psi,\bar{\Psi},\partial_i\bar{\Psi},\Pi,
\bar{\Pi},\mathbf{r},t)\equiv \bar{\Pi}(\partial_0\Psi)+
(\partial_0\bar{\Psi})\Pi-
\mathscr{L}^{(R)}(\Psi,\partial_{\mu}\Psi,\bar{\Psi},\partial_{\mu}\bar{\Psi},\mathbf{r},t)
\end{equation}
where $\bar{\Pi}(\mathbf{r},t)\equiv\partial \mathscr{L}^{(R)}/\partial (\partial_0\Psi)$ and $\Pi(\mathbf{r},t)\equiv\partial \mathscr{L}^{(R)}/\partial (\partial_0\bar{\Psi})$ are the fields canonically conjugated to 
$\Psi(\mathbf{r},t)$ and $\bar{\Psi}(\mathbf{r},t)$ respectively. 
In our case, it can easily be shown that, apart from total derivative terms the Hamiltonian 
density can be written in the form
\begin{equation}
\label{H}
\begin{split}
\mathscr{H}^{(R)}(\Psi,\partial_i\Psi,\bar{\Psi},\mathbf{r},t) &=
64\Psi^{\dag}(\mathbf{r},t)
\mathcal{H}^{(R)}(\mathbf{r},-i\boldsymbol{\partial},t)\Psi(\mathbf{r},t)
\end{split}
\end{equation}
where
\begin{equation}
\label{H_1p}
\mathcal{H}^{(R)}(\mathbf{r},-i\boldsymbol{\partial},t)=\frac{2z}{r_G}
\left\{\frac{\alpha_x}{4}[-i\partial_x+eA_x(y,t)]+
\frac{\alpha_y}{4}[-i\partial_y+eA_y(x,t)]+\beta m\right\}
-i\frac{\alpha_z}{4}\frac{z\partial_z+\partial_z z}{r_G}
\end{equation}
with $\beta=\gamma^0$, $\alpha_x=\gamma^0\gamma^1$, 
$\alpha_y=\gamma^0\gamma^2$ and 
$\alpha_z=\gamma^0\gamma^3$ is 
the one-particle Hamiltonian of an electron in the presence of the magnetic field 
(\ref{B}) in the spacetime with the metric tensor (\ref{g_R}).

If we define, now, 
the scalar product between two generic spinors $\psi_1(\mathbf{r},t)$ and 
$\psi_2(\mathbf{r},t)$ as
\begin{equation}
\label{s_p}
(\psi_1,\psi_2)\equiv\int_{S}dS_{\mu}\sqrt{-g^{(R)}(z)}\bar{\psi}_1(\mathbf{r},t)
\gamma^{(R)\mu}(z)\psi_2(\mathbf{r},t)
\end{equation}
with $S$ the hyper-surface at constant time then, since 
$\sqrt{-g^{(R)}(z)}=128z/r_G$,
\begin{equation}
\label{s_p_f}
(\psi_1,\psi_2)=\int d\mathbf{r} \frac{128z}{r_G}
\psi^{\dag}_1(\mathbf{r},t)\gamma^0\gamma^0\frac{r_G}{2z}\psi_2(\mathbf{r},t)=
64\int d\mathbf{r} \psi^{\dag}_1(\mathbf{r},t)\psi_2(\mathbf{r},t)
\end{equation}
and the one-particle Hamiltonian 
$\mathcal{H}^{(R)}(\mathbf{r},-i\boldsymbol{\partial},t)$ is Hermitian. This 
definition of the scalar product clarifies the presence of the numerical coefficient 
$64$ in Eq. (\ref{H}). We want also to point out here a fact concerning the 
limits of the integrals on the variables $x$, $y$ and $z$. In fact, in our model 
$|x|\ll r_G$, $|y|\ll r_G$ and $z\ll r_G$, but in what follows 
we will consider only electron and positron wave functions 
that, as $|x|$, $|y|$ or $z$ go to infinity, go to zero exponentially 
with a typical length at most of the order of $\lambdabar$ 
then we can assume the integrals 
on $x$ and $y$ as going from $-\infty$ to $\infty$ and that on $z$ from 
$0$ to $\infty$ without appreciable error.

Finally, the total Hamiltonian of the system under study is
\begin{equation}
\label{H_tot}
H^{(R)}(t)\equiv\int d\mathbf{r}\mathscr{H}^{(R)}
(\Psi,\partial_i\Psi,\bar{\Psi},\mathbf{r},t)=
64\int d\mathbf{r}\Psi^{\dag}(\mathbf{r},t)
\mathcal{H}^{(R)}(\mathbf{r},-i\boldsymbol{\partial},t)\Psi(\mathbf{r},t)
\end{equation}
and it depends explicitly on time through the time dependence of the magnetic field. Now, our next step is the determination of the electron and 
positron one-particle instantaneous eigenstates of the 
one-particle Hamiltonian (\ref{H_1p}).
%
%
\section{Determination of the one-particle states}
\label{III}
\setcounter{equation}{0}
\renewcommand{\theequation}{III.\arabic{equation}}
In this Section we assume that the magnetic field has the same form as in 
Eq. (\ref{B}) but that it does not depend on time. All the quantities, 
such as the one-particle Hamiltonian (\ref{H_1p}), that depended on time through 
the magnetic field will be indicated here with the same symbol 
used in the previous Section but, 
of course, omitting the time-dependence.

If the magnetic field does not depend on time the eigenvalue equations 
\begin{align}
\label{Eig_E_p}
\mathcal{H}^{(R)}(\mathbf{r},-i\boldsymbol{\partial})u_{\jmath}&=w_{\jmath}u_{\jmath},\\
\mathcal{H}^{(R)}(\mathbf{r},-i\boldsymbol{\partial})v_{\jmath}&=-\tilde{w}_{\jmath}v_{\jmath}
\end{align}
with $w_{\jmath},\tilde{w}_{\jmath}>0$ can be solved exactly \cite{Torres,Bautista}. Since in these papers a clear derivation is presented we will give directly the final form of the eigenstates.\footnote{In \cite{Torres} another electromagnetic gauge is used but the calculations can be adapted straightforwardly to our case.} Now, unlike the Minkowski spacetime case, in the Rindler spacetime it is convenient to label the states directly with the energy which is a continuous nonnegative quantum number, that will be indicated as $E$, independent of the others that are the usual nonnegative integers $n_d$ and $n_g$ and the polarization $\sigma=\pm 1$. The fact that in 
the Rindler metric the energy $E$ of the electron has  
continuous eigenvalues from zero to infinity that do not depend on the other quantum numbers is the most relevant difference with respect to the Minkowski spacetime case \cite{Cohen}. The physical origin of this difference lies on the fact that in the present case, due to the presence of the \emph{negative} gravitational potential, the electron mass is no more the lowest energy with which the electron or the positron can be created. We will see in Section 4 how this fact will change also qualitatively the final results with respect to the analogous ones in Minkowski spacetime. 

Now, without going into the details we only quote that in the present case the electron and positron eigenstates can be written respectively as
\begin{equation}
\label{u}
\begin{split}
u_{n_d,n_g,\sigma}(E;\mathbf{r})=\sqrt{\frac{k_{n_d}r_G\cosh(2\pi Er_G)}{4\pi^2}}
&[P_-K_{1/2+2iEr_G}(4k_{n_d}z)+\\
&+P_+K_{1/2-2iEr_G}(4k_{n_d}z)]
\varphi_{n_d,n_g,\sigma}(x,y)
\end{split}
\end{equation}
and
\begin{equation}
\label{v}
\begin{split}
v_{n_d,n_g,\sigma}(E;\mathbf{r})=\sqrt{\frac{k_{n_g}r_G\cosh(2\pi Er_G)}{4\pi^2}}
&[P_-K_{1/2-2iEr_G}(4k_{n_g}z)+\\
&+P_+K_{1/2+2iEr_G}(4k_{n_g}z)]
\tilde{\varphi}_{n_d,n_g,\sigma}(x,y)
\end{split}
\end{equation}
In the previous equations $K_w(x)$ are the modified Bessel functions with complex index regular in the limit $x\to \infty$ \cite{Abramowitz} and $P_{\pm}=(1\pm\alpha_z)/2$. Also, the quantity
\begin{equation}
\label{k}
k_n=\sqrt{m^2+\frac{eBn}{8}}
\end{equation}
can be interpreted as a sort of ``transverse'' energy of the electron 
in the spacetime with the metric (\ref{g_R}) while $\varphi_{n_d,n_g,\sigma}(x,y)$ and $\tilde{\varphi}_{n_d,n_g,\sigma}(x,y)$ are defined as
\begin{equation}
\label{phi}
\varphi_{n_d,n_g,\sigma}(x,y)=\frac{1}{2\sqrt{k_{n_d}}}
\left(\begin{array}{c}
\sqrt{k_{n_d}+m\sigma}\theta_{n_d-1,n_g}(x,y)\\
\sqrt{k_{n_d}-m\sigma}\theta_{n_d,n_g}(x,y)\\
i\sigma\sqrt{k_{n_d}+m\sigma}\theta_{n_d-1,n_g}(x,y)\\
i\sigma\sqrt{k_{n_d}-m\sigma}\theta_{n_d,n_g}(x,y)
\end{array}\right)
\end{equation}
and as
\begin{equation}
\label{tr_p}
\tilde{\varphi}_{n_d,n_g,\sigma}(x,y)=\frac{1}{2\sqrt{k_{n_g}}}\left(\begin{array}{c}
i\sigma\sqrt{k_{n_g}-m\sigma}\theta_{n_g-1,n_d}(x,y)\\
i\sigma\sqrt{k_{n_g}+m\sigma}\theta_{n_g,n_d}(x,y)\\
\sqrt{k_{n_g}-m\sigma}\theta_{n_g-1,n_d}(x,y)\\
\sqrt{k_{n_g}+m\sigma}\theta_{n_g,n_d}(x,y)
\end{array}\right)
\end{equation}
with the functions $\theta_{n_d,n_g}(x,y)$ given by
\begin{equation}
\theta_{n_d,n_g}(x,y)=\sqrt{\frac{eB}{2\pi}\frac{1}{n_d!}\frac{1}{n_g!}}
(a_d^{\dag})^{n_d}(a_g^{\dag})^{n_g}\exp\left[-\frac{eB(x^2+y^2)}{4}\right]
\end{equation}
and $a_d^{\dag}$ and $a_g^{\dag}$ the rising operators relative to the quantum numbers $n_d$ and $n_g$ respectively. It can be shown that the states $u_{n_d,n_g,\sigma}(E;\mathbf{r})$ and $v_{n_d,n_g,\sigma}(E;\mathbf{r})$ are such that [see Eq. (\ref{s_p_f})]
\begin{align}
&(u_{n_d,n_g,\sigma}(E),u_{n'_d,n'_g,\sigma'}(E'))=(v_{n_d,n_g,\sigma}(E),v_{n'_d,n'_g,\sigma'}(E'))=
\delta(E-E')\delta_{n_d,n'_d}\delta_{n_g,n'_g}
\delta_{\sigma,\sigma'},\\
&(u_{n_d,n_g,\sigma}(E),v_{n'_d,n'_g,\sigma'}(E'))=0.
\end{align}
and that the basis $\{u_{n_d,n_g,\sigma}(E;\mathbf{r}),v_{n_d,n_g,\sigma}(E;\mathbf{r})\}$ is complete.

As usual, it is preferable to deal with normalizable 
wave functions then we have to 
find a convenient boundary condition at a given surface $z=b$ that 
discretizes the energies $E$. Since 
the procedure is identical for electron and positron states we will 
consider only the electron states. The functions $K_{1/2\pm 2iEr_G}(4k_{n_d}z)$ go exponentially to 
zero for large values of $k_{n_d}z$ and go to infinity as 
$\left(k_{n_d}z\right)^{-1/2}$ for small 
values of $k_{n_d}z$ \cite{Abramowitz}. For this reason, it is clear 
that 
\begin{enumerate}
\item it is impossible to satisfy a ``zero'' condition 
for the eigenstates $u_{n_d,n_g,\sigma}(E;\mathbf{r}),$ 
at a given $4k_{n_d}b\ll 1$ or a canonical periodicity 
condition between two points $4k_{n_d}b_1\ll 1$ and $4k_{n_d}b_2\gg 1$;
\item if we want to build eigenstates with a finite normalization integral 
we have to modify 
the functions $K_{1/2\pm 2iEr_G}(4k_{n_d}z)$ in the region with 
$k_{n_d}z\ll 1$.
\end{enumerate}
In order to do this we proceed as follows. We consider an arbitrary fixed 
surface $z=b$ such that $k_{n_d}b\ll 1$ and assume that the positive-energy 
eigenstates of the one-particle Hamiltonian  are the spinors $u_{n,n_d,n_g,\sigma}(\mathbf{r})$ defined as
\begin{equation}
\label{u_d}
u_{n,n_d,n_g,\sigma}(\mathbf{r})=
\begin{cases}
\begin{aligned}
N^{(<)}_{n,n_d,n_g,\sigma} &\sqrt{\frac{k_{n_d}r_G
\cosh(2\pi E_{n,n_d}r_G)}{4\pi^2}}\times\\
&\times\left[P_-I_{1/2+2iE_{n,n_d}r_G}(4k_{n_d}z)+\right.\\
&\quad\left.+P_+I_{1/2-2iE_{n,n_d}r_G}(4k_{n_d}z)\right]
\varphi_{n_d,n_g,\sigma}(x,y) & \text{if $z\le b$}
\end{aligned}\\
\begin{aligned}
N^{(>)}_{n,n_d,n_g,\sigma} &\sqrt{\frac{k_{n_d}r_G
\cosh(2\pi E_{n,n_d}r_G)}{4\pi^2}}\times\\
&\times\left[P_-K_{1/2+2iE_{n,n_d}r_G}(4k_{n_d}z)+\right.\\
&\quad\left.+P_+K_{1/2-2iE_{n,n_d}r_G}(4k_{n_d}z)\right]
\varphi_{n_d,n_g,\sigma}(x,y) & \text{if $z> b$}
\end{aligned}
\end{cases}
\end{equation}
where $I_w(x)$ are the modified Bessel functions with complex index regular at $x=0$ \cite{Abramowitz}, $n$ is a new 
integer quantum number characterizing the discrete 
energies (as we will see these energies will also depend on the quantum number 
$n_d$) and $N^{(<)}_{n,n_d,n_g,\sigma}$ and $N^{(>)}_{n,n_d,n_g,\sigma}$ are two real 
normalization factors to be determined. In fact, by solving step by step the eigenvalue equations (\ref{Eig_E_p}) one sees that they can be disentangled into an equation depending on $x$ and $y$ and another one depending only on $z$ which is indeed a modified Bessel equation. The solution space of this last equation is spanned by the functions $I_{1/2\pm 2iE_{n,n_d}r_G}(4k_{n_d}z)$ and $K_{1/2\pm 2iE_{n,n_d}r_G}(4k_{n_d}z)$ and we have chosen the $K_{1/2\pm 2iE_{n,n_d}r_G}(4k_{n_d}z)$ because, unlike the $I_{1/2\pm 2iE_{n,n_d}r_G}(4k_{n_d}z)$, they are regular in the limit $z\to \infty$. Now, since Eq. (\ref{Eig_E_p}) is a first-order 
equation in the variable $z$ we only require that the 
spinor $u_{n,n_d,n_g,\sigma}(\mathbf{r})$ is continuous at $z=b$. 
By means of this condition and by requiring that the norm of 
$u_{n,n_d,n_g,\sigma}(\mathbf{r})$ is the unit, we make the energies discrete and determine 
the normalization factors $N^{(<)}_{n,n_d,n_g,\sigma}$ and 
$N^{(>)}_{n,n_d,n_g,\sigma}$. The details of the 
calculations are given in the appendix and here we only quote the final expression of the 
discrete energies [see Eq. (\ref{E_n_n_d})]
\begin{align}
E_{n,n_d}r_G\log \left(k_{n_d}b\right)=n\frac{\pi}{2} && k_{n_d}b\to 0
\end{align}
and of the coefficients $N^{(<)}_{n,n_d,n_g,\sigma}$ and 
$N^{(>)}_{n,n_d,n_g,\sigma}$ [see Eqs. (\ref{N_m}) and (\ref{N_M})]:
\begin{align}
\label{N_m_t}
N^{(<)}_{n,n_d,n_g,\sigma} &=N^{(<)}_{n,n_d}=
\frac{\pi}{8k_{n_d}b}\frac{\sqrt{1+(4E_{n,n_d}r_G)^2}}
{\cosh(2\pi E_{n,n_d}r_G)}\frac{1}{\sqrt{\varrho_{n_d}}} && k_{n_d}b\to 0,\\
\label{N_M_t}
N^{(>)}_{n,n_d,n_g,\sigma} &=N^{(>)}_{n_d}=\frac{1}{\sqrt{\varrho_{n_d}}} 
 && k_{n_d}b\to 0
\end{align}
with [see Eq. (\ref{rho})]
\begin{align}
\label{rho_t}
\varrho_{n_d}=-\frac{2r_G}{\pi}\log(k_{n_d}b) && k_{n_d}b\to 0
\end{align}
the density of the energy levels. Obviously, all these 
quantities will be used in the calculations but at the end we have to perform 
the limit $b\to 0$ and the physically relevant results 
must be independent of $b$.
%
%
\section{Calculation of the production probability}
\label{IV}
\setcounter{equation}{0}
\renewcommand{\theequation}{IV.\arabic{equation}}
In the framework of the adiabatic perturbation theory, 
the matrix element of the creation process of a pair with the electron 
in the state $u_J(\mathbf{r},t)$ with $J\equiv\{n,n_d,n_g,\sigma\}$ and the positron in the state 
$v_{J'}(\mathbf{r},t)$ with $J'\equiv\{n',n'_d,n'_g,\sigma'\}$ is given by \cite{Migdal}
\begin{equation}
\label{M_E_i}
\dot{H}^{(R)}_{JJ'}(t)\equiv\langle 1_J(t);
\tilde{1}_{J'}(t)|\dot{H}^{(R)}(t)|0(t)\rangle
=\frac{16e\dot{B}(t)}{r_G}\int d\mathbf{r}u_J^{\dag}(\mathbf{r},t)
z\left(x\alpha_y-y\alpha_x\right)v_{J'}(\mathbf{r},t)
\end{equation}
where $|0(t)\rangle$ and 
$|1_J(t);\tilde{1}_{J'}(t)\rangle
\equiv c^{\dag}_J(t)d^{\dag}_{J'}(t)|0(t)\rangle$ 
are the vacuum and the pair state at time $t$ respectively 
[see Eqs. (\ref{H_tot}), (\ref{H}) and (\ref{H_1p})]. The factor 
$\dot{B}(t)(x\alpha_y-y\alpha_x)$, using Eqs. (\ref{A_1}) and (\ref{A_2}) and the
expression of the induced electric field 
$\mathbf{E}(\mathbf{r},t)=-\partial\mathbf{A}(\mathbf{r},t)/\partial t$, can be
rewritten in terms of the scalar product of the electric field and the velocity 
operator $\boldsymbol{\alpha}$ that is of the work per unit
time done by the induced electric field itself. We incidentally observe that the electric field $\mathbf{E}(\mathbf{r},t)$ is always perpendicular to the magnetic field $\mathbf{B}(t)$. A more useful form of the previous 
matrix element can be given by using 
the matrices $\alpha_{\pm}=(\alpha_x\pm i\alpha_y)/2$ and the relations
\begin{align}
x &=\frac{1}{2}\sqrt{\frac{2}{eB(t)}}[a_g(t)+a_g^{\dag}(t)+a_d(t)+a_d^{\dag}(t)],\\
y &=\frac{1}{2i}\sqrt{\frac{2}{eB(t)}}[a_g(t)-a_g^{\dag}(t)-a_d(t)+a_d^{\dag}(t)]
\end{align}
with $a_d(t)$ [$a^{\dag}_d(t)$] and $a_g(t)$ [$a^{\dag}_g(t)$] the 
lowering [rising] operators relative to the quantum numbers $n_d$ 
and $n_g$ respectively. The result is
\begin{equation}
\label{M_E}
\dot{H}^{(R)}_{JJ'}(t)=\frac{32ie\dot{B}(t)}{r_G\sqrt{2eB(t)}}
\int d\mathbf{r}u_J^{\dag}(\mathbf{r},t)
z\{\alpha_-[a_d(t)+a_g^{\dag}(t)]-
\alpha_+[a_g(t)+a_d^{\dag}(t)]\}v_{J'}(\mathbf{r},t).
\end{equation}

Now, in the rest of this Section we will first manipulate the matrix element (\ref{M_E}) to put it in the form (\ref{M_E_2}). Then we will use it to calculate the production probabilities [Eqs. (\ref{dP_1}) and (\ref{dP_2})] by means of the usual adiabatic perturbation theory and finally the total production probability per unit volume and unit energy (\ref{dP_f_f}).

As we have said at the end of the previous Section, 
in order to calculate this matrix elements we should 
use the expression (\ref{u_d}) for $u_J(\mathbf{r},t)$ with 
$N^{(<)}_{n,n_d}(t)$ and $N^{(>)}_{n_d}(t)$ given 
by Eqs. (\ref{N_m_t}) and (\ref{N_M_t}) respectively 
and the analogous expression for $v_J^{\dag}(\mathbf{r},t)$
\footnote{Remind that these quantities are now time-depending because
the magnetic field depends on time.}. Actually, an easy power 
counting will show that the contribution of the integral on the variable $z$ from 
$0$ to $b$ goes to $0$ in the limit $b\to 0$. In fact, each spinor contains a factor 
$1/\big[k_{n_d}(t)b\sqrt{\log(k_{n_d}(t)b)}\big]$ 
coming from $N^{(<)}_{n,n_d}(t)$ 
[see Eqs. (\ref{N_m_t}) and (\ref{rho_t})]. Also, from Eq. (\ref{I_a}) we see that 
the modified Bessel functions $I_{1/2+2iE_{n,n_d}(t)r_G}(4k_{n_d}(t)z)$ 
behave as $\sqrt{k_{n_d}(t)z}$ in the integration 
domain $0\le z\le b$. Finally, because of the presence of the $z$ factor in 
the matrix element (\ref{M_E}) 
the result of the integral on $z$ depends on $b$ as 
$k_{n_d}(t)b/\log(k_{n_d}(t)b)$ and 
then it goes to zero in the limit $b\to 0$. In this way, since at the end 
of the calculations the limit $b\to 0$ has to be performed, the matrix element 
(\ref{M_E}) can be calculated by using in the whole region 
$z\ge 0$ the expressions of the spinors $u_J(\mathbf{r},t)$ and 
$v_{J'}(\mathbf{r},t)$ valid in the region $z>b$. Actually, we can use 
directly the spinors $u_{n_d,n_g,\sigma}(E;\mathbf{r})$ and 
$v_{n'_d,n'_g,\sigma'}(E';\mathbf{r})$ multiplied by $N^{(>)}_{n_d}(t)$ and 
$N^{(>)}_{n'_g}(t)$ respectively because the presence of the factor $z$ in 
the matrix element (\ref{M_E}) makes finite the resulting integral 
from $0$ to $\infty$ [see also the general 
formula (\ref{int})]:
\begin{equation}
\label{M_E_1}
\begin{split}
\dot{H}^{(R)}_{n_d,n_g,\sigma;n'_d,n'_g,\sigma'}(E,E';t)= 
&\frac{32ie\dot{B}(t)}{r_G\sqrt{2eB(t)\varrho_{n_d}(t)\varrho_{n'_g}(t)}}\times\\
&\times\int d\mathbf{r}u_{n_d,n_g,\sigma}^{\dag}(E;\mathbf{r},t)
z\{\alpha_-[a_d(t)+a_g^{\dag}(t)]-\\
&\qquad\qquad\qquad\qquad\qquad\quad-\alpha_+[a_g(t)+a_d^{\dag}(t)]\}
v_{n'_d,n'_g,\sigma'}(E';\mathbf{r},t).
\end{split}
\end{equation}
At this point we have to substitute Eqs. (\ref{u}) and (\ref{v}) 
with the time dependent magnetic field in place 
of $u_{n_d,n_g,\sigma}^{\dag}(E;\mathbf{r},t)$ and 
$v_{n'_d,n'_g,\sigma'}(E';\mathbf{r},t)$ and apply the various operators. The calculations are involved but straightforward. We will not report them here but we give the final result
\begin{equation}
\label{M_E_2}
\begin{split}
\dot{H}^{(R)}_{n_d,n_g,\sigma;n'_d,n'_g,\sigma'}(E,E';t) 
&=\frac{ie\dot{B}(t)}{\pi^2}
\sqrt{\frac{\cosh(2\pi Er_G)\cosh(2\pi E'r_G)}{2[eB(t)]^3}}\times\\
&\times\left\{\sqrt{\frac{(n_g+1)\left[k_{n_d}(t)-m\sigma\right]
\left[k_{n_d+1}(t)-m\sigma'\right]}{\varrho_{n_d}(t)\varrho_{n_d+1}(t)}}\right.\times\\
&\qquad\times\mathrm{Re}\left((1-i\sigma)(1+i\sigma')\mathcal{I}_{n_d,n_d+1}(E,E';t)\right)
\delta_{n_d,n'_g-1}\delta_{n_g+1,n'_d}+\\
&\quad+\sqrt{\frac{n_d\left[k_{n_d}(t)-m\sigma\right]
\left[k_{n_d}(t)-m\sigma'\right]}{\varrho^2_{n_d}(t)}}\times\\
&\qquad\times\mathrm{Re}\left((1-i\sigma)(1+i\sigma')\mathcal{I}_{n_d,n_d}(E,E';t)\right)
\delta_{n_d,n'_g}\delta_{n_g,n'_d}-\\
&\quad-\sqrt{\frac{n_g\left[k_{n_d}(t)+m\sigma\right]
\left[k_{n_d-1}(t)+m\sigma'\right]}{\varrho_{n_d}(t)\varrho_{n_d-1}(t)}}\times\\
&\qquad\times\mathrm{Re}\left((1+i\sigma)(1-i\sigma')\mathcal{I}_{n_d,n_d-1}(E,E';t)\right)
\delta_{n_d-1,n'_g}\delta_{n_g-1,n'_d}-\\
&\quad-\sqrt{\frac{n_d\left[k_{n_d}(t)+m\sigma\right]
\left[k_{n_d}(t)+m\sigma'\right]}{\varrho^2_{n_d}(t)}}\times\\
&\qquad\times\mathrm{Re}\left((1+i\sigma)(1-i\sigma')\mathcal{I}_{n_d,n_d}(E,E';t)\right)
\delta_{n_d,n'_g}\delta_{n_g,n'_d}\Biggr\}
\end{split}
\end{equation}
where the adimensional function
\begin{equation}
\label{inte}
\mathcal{I}_{l,l'}(E,E';t)\equiv\int_0^{\infty}ds s
K_{1/2-2iEr_G}\left(\frac{4k_l(t)}{\sqrt{2eB(t)}}s\right)
K_{1/2+2iE'r_G}\left(\frac{4k_{l'}(t)}{\sqrt{2eB(t)}}s\right)
\end{equation}
has been introduced.

Before continuing we want to point out that from Eq. (\ref{M_E_2}) it can be 
seen that the total angular momentum of the electron-positron field is conserved in the 
transition in fact, in any case
\begin{equation}
n_d-n_g-\frac{1}{2}+n'_d-n'_g+\frac{1}{2}=n_d-n_g+n'_d-n'_g=0.
\end{equation}
Of course, this selection rule is a consequence of the fact that 
the time evolution of the magnetic field does not break the rotational symmetry of the 
system around the $z$ axis or, in other words, of the fact that 
the total angular momentum along $z$ and 
$\dot{\mathcal{H}}^{(R)}(\mathbf{r},-i\boldsymbol{\partial},t)$ commute.

We also observe that, 
as in the Minkowski spacetime, it is impossible in the case under study 
to create a pair in which the electron is in a $(n_d=0,\sigma=-1)$-state 
and the positron in a $(n_g=0,\sigma=+1)$-state 
\cite{Calucci,DiPiazza2}. In fact, we remind that 
this selection rule holds in general for the eigenstates of 
$\sigma_z$ when $\sigma_z$ anticommute with the time-derivative of the 
one-particle Hamiltonian \cite{DiPiazza2} and it can easily be shown that 
this is true in our present case because the gravitational field changes only 
the longitudinal structure of the one-particle electron and positron 
wave functions.

Now, since we are interested only in the strong magnetic field regime 
in which $eB(t)\gg m^2$, we can simplify the expression of the 
transition matrix element (\ref{M_E_2}) by taking into account only those 
transitions whose probabilities are proportional to the 
lowest power of $m^2/eB(t)$. In the framework of the 
adiabatic perturbation theory the first-order transition amplitude in $\dot{B}(t)$ 
of the creation of a pair at time $t$ in the state with quantum numbers
$\{E,n_d,n_g,\sigma;E',n'_d,n'_g,\sigma'\}$ is given by \cite{Migdal}
\begin{equation}
\label{ampl}
\gamma_{n_d,n_g,\sigma;n'_d,n'_g,\sigma'}(E,E';t)=
\frac{1}{E+E'}\int_0^tdt'
\dot{H}^{(R)}_{n_d,n_g,\sigma;n'_d,n'_g,\sigma'}(E,E';t')\exp{(i(E+E')t')}
\end{equation}
and the corresponding probability is the square modulus of this number. It 
is evident that, since the energies $E$ do not depend on $B(t)$, we can perform 
the $(m^2/eB(t))$-power counting directly on the matrix element (\ref{M_E_2}). 
To this end we need the general 
behaviour of two particular classes of the integral (\ref{inte}) that is 
$\mathcal{I}_{0,n'_d}(E,E';t)$ with $n'_d>0$ and $\mathcal{I}_{n_d,n_d}(E,E';t)$ with 
$n_d>0$. By reminding the expression (\ref{k}) with the time-dependent 
magnetic field for $k_{n_d}(t)$ and by using the general formula (\ref{int}) it 
can easily be seen that
\begin{align}
\mathcal{I}_{0,n'_d}(E,E';t) &\sim \frac{1}{\sqrt{n^{\prime 3}_d}}
\left[\frac{m}{\sqrt{eB(t)}}\right]^{1/2} 
&& \text{if $n'_d>0$ and $eB(t)\gg m^2$},\\
\label{I_nn}
\mathcal{I}_{n_d,n_d}(E,E';t) &\sim \frac{1}{n_d} 
&& \text{if $n_d>0$ and $eB(t)\gg m^2$}
\end{align}
where, for later convenience, we have also pointed out the dependence on the quantum 
numbers $n_d$ and $n'_d$. Obviously, the integrals 
$\mathcal{I}_{n_d,n_d\pm 1}(E,E';t)$ behave as the integral (\ref{I_nn}) and then 
this criterion allows us to neglect the transitions 
in which the electron is in a $(n_d=0,\sigma=-1)$-state 
or the positron in a $(n_g=0,\sigma=+1)$-state [see Eq. (\ref{M_E_2})].

Another criterion we will use to select only the most probable transitions 
is the dependence of the corresponding probabilities 
on the quantum numbers $n_d$ and $n_g$. As previously, we can work directly on 
the matrix element (\ref{M_E_2}) by keeping in mind that at the end 
we will sum the probabilities with different values of $n_d$ and $n_g$. Now, 
we have explained in \cite{DiPiazza2} that the internal consistency of the model 
requires that the sum on $n_g$ (and on $n_d$) 
cannot be extended up to infinity but that they must be stopped 
up to a certain $N_M(t)$ corresponding through the 
relation [see also \cite{DiPiazza2}]
\begin{equation}
\label{N_max}
N_M(t)\equiv \frac{eB(t)}{32}R^2_{\perp M}
\end{equation}
to a fixed $R_{\perp M}$ whose physical meaning is explained below (the presence of the factor $1/32$ is due to the fact that the spatial metric in Eq. (\ref{g_R}) is only proportional to the Euclidean one). Now, since we are considering the production of electrons (positrons) in states also with $n_d$ ($n_g$) different from zero, in general the transverse motion of a classical electron (positron) is confined within a circle with radius $2R_{\perp M}$, then this quantity can be assumed as the radius of the quantization cylinder. Coming back to 
the matrix element (\ref{M_E_2}), we see that it contains 
two kinds of terms, the first one being proportional 
essentially to $\sqrt{n_g}$ and the second one to $\sqrt{n_d}$. By taking into account 
Eq. (\ref{I_nn}) it is easy to see that the first kind of terms gives rise 
to final probabilities proportional to $N_M^2(t)\log N_M(t)$, while the 
second one to final probabilities proportional to $N_M^2(t)$. For all these 
reasons we can consider only the transitions to states with 
$n_d> 0$ and $n'_g> 0$ and approximate the matrix element (\ref{M_E_2}) as
\begin{equation}
\label{M_E_3}
\begin{split}
\dot{H}^{(R)}_{n_d,n_g,\sigma;n'_g,n'_d,\sigma'}(E,E';t) 
&\simeq\frac{i\dot{B}(t)}{(2\pi)^2B(t)}
\sqrt{\cosh(2\pi Er_G)\cosh(2\pi E'r_G)}\times\\
&\times\Biggr[\sqrt{\frac{(n_g+1)\sqrt{n_d(n_d+1)}}
{\varrho_{n_d}(t)\varrho_{n_d+1}(t)}}\times\\
&\qquad\times\mathrm{Re}\left((1-i\sigma)(1+i\sigma')\mathcal{I}_{n_d,n_d+1}(E,E')\right)
\delta_{n_d,n'_g-1}\delta_{n_g+1,n'_d}+\\
&\quad-\sqrt{\frac{n_g\sqrt{n_d(n_d-1)}}{\varrho_{n_d}(t)\varrho_{n_d-1}(t)}}\times\\
&\qquad\times
\mathrm{Re}\left((1+i\sigma)(1-i\sigma')\mathcal{I}_{n_d,n_d-1}(E,E')\right)
\delta_{n_d-1,n'_g}\delta_{n_g-1,n'_d}\Biggr]
\end{split}
\end{equation}
where we have pointed out that in the strong magnetic field regime 
if $n_d> 0$ and $n'_g> 0$ the integrals 
$\mathcal{I}_{n_d,n_d\pm 1}(E,E')$ do not depend on time 
[see Eq. (\ref{int})]. 

At this point, by inserting this matrix element in Eq. (\ref{ampl}), by squaring and summing on the polarization variable $\sigma$ we 
obtain the differential probabilities
\begin{align}
\label{dP_1}
\begin{split}
dP_{n_d,n_g;n_d+1,n_g+1}(E,E';t) &=
\frac{1}{2\pi^4}(n_g+1)\sqrt{n_d(n_d+1)}\cosh(2\pi Er_G)\cosh(2\pi E'r_G)\times\\
&\times\frac{\left\vert\mathcal{I}_{n_d,n_d+1}(E,E')\right\vert^2}{(E+E')^2}
\left\vert\mathcal{F}(E,E';t)\right\vert^2dEdE',
\end{split}\\
\label{dP_2}
\begin{split}
dP_{n_d+1,n_g+1;n_d,n_g}(E,E';t) &=
\frac{1}{2\pi^4}(n_g+1)\sqrt{n_d(n_d+1)}\cosh(2\pi Er_G)\cosh(2\pi E'r_G)\times\\
&\times\frac{\left\vert\mathcal{I}_{n_d+1,n_d}(E,E')\right\vert^2}{(E+E')^2}
\left\vert\mathcal{F}(E,E';t)\right\vert^2dEdE'
\end{split}
\end{align}
where we have multiplied by the number of electronic states $\varrho_{n_d}(t)dE$ with energies 
between $E$ and $E+dE$ and by the number of positronic states 
$\varrho_{n_d+1}(t)dE'$ with energies 
between $E'$ and $E'+dE'$ and where
\begin{equation}
\label{F}
\mathcal{F}(E,E';t)\equiv\int_0^tdt'\frac{\dot{B}(t')}{B(t')}\exp{(i(E+E')t')}.
\end{equation}
We point out that, as expected, the probabilities (\ref{dP_1}) and (\ref{dP_2}) do not depend on the unphysical parameter $b$.

Now, we want to calculate the probability $dP(E,E';t)$ 
that a pair is present at time $t$ with the electron with 
energy between $E$ and $E+dE$ and the positron 
with energy between $E'$ and $E'+dE'$. To do this we have to sum on 
the remaining quantum numbers $n_d$ and 
$n_g$. As we already know, both 
the series on $n_d$ and $n_g$ are diverging then we can perform the summations 
by assuming $n_g\simeq n_g+1$ and $n_d\simeq n_d+1$ because the most relevant 
terms are those with $n_g\gg 1$ and $n_d\gg 1$. Starting 
from Eqs. (\ref{dP_1}) and (\ref{dP_2}) we have
\begin{equation}
\label{dP}
\begin{split}
dP(E,E';t)\simeq
\frac{1}{\pi^4}\left[\sum_{n_g=1}^{N_M(t)}n_g\right]
&\cosh(2\pi Er_G)\cosh(2\pi E'r_G)\times\\
&\times\frac{\sum_{n_d=1}^{N_M(t)}n_d
\left\vert\mathcal{I}_{n_d,n_d}(E,E')\right\vert^2}{(E+E')^2}
\left\vert\mathcal{F}(E,E';t)\right\vert^2dEdE'
\end{split}
\end{equation}
where $N_M(t)$ has been defined in Eq. (\ref{N_max}). 

The next step is 
the explicit calculation of the functions 
$\mathcal{I}_{n_d,n_d}(E,E')$ and $\mathcal{F}(E,E';t)$ defined in Eqs. (\ref{inte}) 
and (\ref{F}). By using the general formula (\ref{int}) and the properties 
of the $\Gamma$ function \cite{Abramowitz} it can easily be shown that
\begin{equation}
\label{int_f}
\mathcal{I}_{n_d,n_d}(E,E')=\frac{\pi}{2n_d}
\frac{1-i(E-E')r_G}{\cosh \left(\pi(E-E')r_G\right)}
\frac{\pi(E+E')r_G}{\sinh \left(\pi(E+E')r_G\right)}.
\end{equation}

In order to evaluate the function $\mathcal{F}(E,E';t)$ we have to assign 
the time dependence of the magnetic field. We assume 
that
\begin{equation}
\label{B_t}
B(t)=B_f+(B_i-B_f)\exp\left(-\frac{t}{\tau}\right)
\end{equation}
with $B_i<B_f$ and it results
\begin{equation}
\label{F_1}
\mathcal{F}(E,E';t)=\int_0^{t/\tau}ds'\frac{B_f-B_i}{B_f+(B_i-B_f)\exp{(-s')}}
\exp{\left(-s'+i(E+E')\tau s'\right)}
\end{equation}
with $s'=t'/\tau$. Now, from a physical point of view we are interested 
only in the cases such that $E\tau\gg 1$ and $E'\tau\gg 1$. 
In fact, the eigenvalues $E$ correspond to 
the classical energies \cite{Landau2}
\begin{equation}
E^{(cl)}(z)=\frac{4m}{\sqrt{1-v^2}}\frac{2z}{r_G}
\end{equation}
with $v^2$ the square of physical velocity as measured by 
the local observer in the gravitational field. Also, $\tau$ is a macroscopic 
time parameter connected to the typical evolution time of the black 
hole. Analytic estimates of the time duration of the 
formation of a black hole suggest 
that $\tau\gtrsim r_G$ \cite{Landau2}. In this hypotheses, 
even if $z\sim \lambdabar$ then 
$E^{(cl)}(z)\tau\gg 1$ because we are interested in 
energetic electrons with a Lorentz factor $1/\sqrt{1-v^2}\gg 1$. For this 
reason we can give the following asymptotic estimate of the integral (\ref{F_1})
\begin{equation}
\mathcal{F}(E,E';t)\sim\frac{B_f-B_i}{i(E+E')\tau}
\left[\frac{\exp{\left(-t/\tau+i(E+E')t\right)}}{B_f+(B_i-B_f)\exp{(-t/\tau)}}-
\frac{1}{B_i}\right]
\end{equation}
or, as $t\to\infty$,
\begin{equation}
\mathcal{F}(E,E';t\to\infty)\sim\frac{B_f-B_i}{B_i}\frac{i}{(E+E')\tau}.
\end{equation}
By substituting this expression and Eq. (\ref{int_f}) in Eq. (\ref{dP}) we can write 
the asymptotic value of the probability $dP(E,E';t\to\infty)$ as
\begin{equation}
\label{dP_f}
dP(E,E';t\to\infty) \sim\frac{1}{2}\left(\frac{eB_fR^2_{\perp M}}{64}\right)^2
\log\left(\frac{eB_fR^2_{\perp M}}{32}\right)
\left(\frac{B_f-B_i}{B_i\tau}\right)^2r_G^4\mathcal{G}(r_G;E,E')dEdE'
\end{equation}
where we have made the substitutions [see Eq. (\ref{N_max})]
\begin{align}
\sum_{n_g=1}^{N_M(t\to\infty)} n_g&\longrightarrow \frac{1}{2}N_M^2(t\to\infty)=
\frac{1}{2}\left(\frac{eB_fR^2_{\perp M}}{32}\right)^2,\\
\sum_{n_d=1}^{N_M(t\to\infty)} \frac{1}{n_d}&\longrightarrow 
\log \left(N_M(t\to\infty)\right)=
\log\left(\frac{eB_fR^2_{\perp M}}{32}\right)
\end{align}
and where the function
\begin{equation}
\label{G}
\mathcal{G}(r_G;E,E')=\frac{1+[(E-E')r_G]^2}{[(E+E')r_G]^2}
\frac{\cosh(2\pi Er_G)\cosh(2\pi E'r_G)}
{\cosh^2(\pi (E-E')r_G)\sinh^2(\pi (E+E')r_G)}
\end{equation}
has been introduced. We have pointed out the dependence of 
$\mathcal{G}(r_G;E,E')$ on the parameter $r_G$ because, as 
we have said, we are interested in energies $E$ and $E'$ such that 
$Er_G\gg 1$ and $E'r_G\gg 1$. In this energy region the function 
$\mathcal{G}(r_G;E,E')$ strongly depends even on small 
changes of $E$ and $E'$ through the hyperbolic functions. In particular, it can be seen that 
\begin{equation}
\begin{split}
\mathcal{G}(r_G;E,E') &=\frac{1+[(E-E')r_G]^2}{[(E+E')r_G]^2}
\left[\frac{1}{\sinh^2(\pi (E+E')r_G)}+
\frac{1}{\cosh^2(\pi (E-E')r_G)}\right].
\end{split}
\end{equation}
From this expression and by reminding that $E,E'\ge 0$, one can easily show that
\begin{equation}
\lim_{r_G\to\infty}\frac{r_G[(E+E')r_G]^2}{1+[(E-E')r_G]^2}\mathcal{G}(r_G;E,E')=
\frac{2}{\pi}\delta (E-E')
\end{equation}
and then that
\begin{align}
\mathcal{G}(r_G;E,E')\sim\frac{1}{2\pi(Er_G)^2}\frac{\delta(E-E')}{r_G} && 
\text{if $E,E'\ge 0$ and $Er_G,E'r_G\gg 1$}.
\end{align}
With this result and by integrating Eq. (\ref{dP_f}) 
with respect to the positron energy $E'$ we, finally, obtain 
the probability that an electron is created at time $t\to\infty$ with 
an energy between $E$ and $E+dE$ such that $E\tau\gtrsim Er_G\gg 1$ in the form
\begin{equation}
dP(E;t\to\infty)\sim \frac{1}{4\pi}
\left(\frac{eB_fR^2_{\perp M}}{64}\right)^2
\log\left(\frac{eB_fR^2_{\perp M}}{32}\right)
\left(\frac{B_f-B_i}{\tau B_i}\right)^2\frac{r_GdE}{E^2}.
\end{equation}
In order to obtain a probability per unit volume we have to give an estimate 
of the height of the quantization cylinder. Now, we have said that 
the modified Bessel functions $K_{1/2\pm 2iEr_G}(4k_{n_d}(t)z)$ 
(we refer to the electron wave functions but an identical 
conclusion can be drawn for the positron ones) are 
exponentially decreasing as $k_{n_d}(t)z\gg 1$. Now, by using the well-known oscillation theorems it can be shown that if $Er_G\gg 1$ 
the exponential behaviour of the function $K_{1/2\pm 2iEr_G}(4k_{n_d}(t\to\infty)z)$ starts at
\begin{equation}
z_0\simeq\frac{2Er_G}{4k_{n_d}(t\to\infty)}
\le\frac{2Er_G}{\sqrt{2eB_f}}.
\end{equation}
For this reason we can assume
\begin{equation} 
V=\pi (2R_{\perp M})^2\times \frac{2Er_G}{\sqrt{2eB_f}}=
\frac{8\pi Er_GR_{\perp M}^2}{\sqrt{2eB_f}}
\end{equation}
as the volume of the quantization cylinder and then
\begin{equation}
\label{dP_f_f}
\frac{dP(E;t\to\infty)}{dVdE}\sim \frac{1}{\pi}
\left(\frac{eB_f}{128}\right)^{5/2}
\log\left(\frac{eB_fR^2_{\perp M}}{32}\right)
\left(\frac{B_f-B_i}{\tau B_i}\right)^2\frac{R^2_{\perp M}}{E^3}.
\end{equation}
We first observe that the disappearance of the electron mass and of the gravitational 
radius of the black hole is due 
only to the fact that we are working in the 
strong magnetic field regime and in the high-energy region. In order to 
clarify the meaning of some variables appearing in Eq. (\ref{dP_f_f}) it may
be useful to recall the physical model which is the starting point of the whole
investigation.
There is a bundle of lines of magnetic flux that is limited transversally but its size is very large at the
microscopic scale where the field may be considered uniform. We remark
that the existence of two scales, one of astrophysical origin and one set by the
elementary particles is essential in the whole treatment. The time variation,
slow at microscopic scale, may involve both changes in strength and in
direction of the magnetic field and in the present case, as in \cite{Calucci}, only the change in strength is
considered. In any case the variation of $\mathbf{B}(t)$ gives rise to an electric
field which is not uniform in the transverse variables: starting from the center of
the bundle the electric field produced by the magnetic-flux variation is 
$[\mathbf{r}\times\dot{\mathbf{B}}(t)]/2$, 
therefore the factor $(B_f-B_i)R_{\perp M}/\tau$ gives the 
order of magnitude of the induced electric field at the boundary of the volume. 
Concerning this fact, we point out that a meaningful interpretation of $R_{\perp M}$ 
can be given. In fact, $R_{\perp M}$ can be seen as the typical length scale in which the magnetic 
(gravitational) field produced by the astrophysical compact object can be 
assumed to be uniform.

Also, before comparing Eq. (\ref{dP_f_f}) 
to the analogous result in \cite{Calucci} where no 
gravitational field effects were taken into account we have to 
stress that the comparison can be 
only qualitative because in \cite{Calucci} a different kind of 
magnetic field time-dependence was used. Nevertheless, we note 
in the present case a \emph{qualitatively} completely different dependence on the physical quantities like the magnetic field strength or the electron (positron) energy. In fact, here the probability per unit volume 
depends on the $(5/2)$-power of the magnetic field strength while 
in \cite{Calucci} it depended on the $(3/2)$-power of the magnetic field strength. Also, the probability production per unit energy scales here as $E^{-3}$ in the high-energy region while in the preceding case the probability 
behaved as $E^{-4}$. In this way, a general enhancement of the 
pair production is observed and the production of high-energy 
pairs is strongly favored in the 
presence of the gravitational field then \emph{the effects of the gravitational field in 
the pair production process are really relevant and they 
can not be neglected}. 

To give a quantitative estimate we first integrate Eq. (\ref{dP_f_f}) from $E_m=100\;r_G^{-1}$ (remind that we assumed $Er_G\gg 1$) to infinity. By indicating the resulting total probability per unit volume as $dP(t\to \infty)/dV$, we obtain
\begin{equation}
\frac{dP(t\to \infty)}{dV}\sim \frac{1}{\pi}
\left(\frac{eB_f}{64}\right)^{5/2}
\log\left(\frac{eB_fR^2_{\perp M}}{32}\right)
\left(\frac{B_f-B_i}{\tau B_i}\right)^2\frac{R^2_{\perp M}}{E_m^2}.
\end{equation}
Now, by assuming $r_G=4.4\;\text{Km}$ as for a $1.5$ solar masses black hole and $\tau=10^{-4}\;\text{s}\gtrsim r_G$, it also results $E_m\tau\gg 1$. By using the typical values $R_{\perp M}=r_G$, $B_i=B_f/10=10^{14}\;\text{G}$ and by assuming the magnetic field parameters used in \cite{Calucci} to be such that $B_0=B_i$, $B(t)\sim B_f$ and $b=(B_f-B_i)/\tau$ we obtain that the ratio between Eq. (20) in \cite{Calucci} (divided by the volume $Z\pi R^2_{\perp M}$) and the previous probability is of the order of $10^{-28}$ that clearly shows how relevant are the effects of the presence of the gravitational field in 
the pair production process. 

Finally, by using the formula
\begin{equation}
\label{dN_f_f}
\mathcal{E}=\int dVdE\;2E\frac{dP(E;t\to\infty)}{dVdE}\sim \frac{2}{\pi}
\left(\frac{eB_f}{128}\right)^{5/2}
\log\left(\frac{eB_fR^2_{\perp M}}{32}\right)
\left(\frac{B_f-B_i}{\tau B_i}\right)^2\frac{R^2_{\perp M}V}{E_m}
\end{equation}
we give an estimate of the total energy produced through the mechanism at hand. By substituting the previous numerical values and $V^{1/3}=r_G$ we obtain $\mathcal{E}\sim 10^{57}\;\text{erg}$. This result must be considered with some care because it is roughly two orders of magnitude larger than the rest energy of a black hole with a mass $1.5$ times the solar mass (on the other hand, the magnetic energy associated to the magnetic field is negligible with respect to the black hole rest energy). One reason of this overestimate is that for simplicity we have assumed the pair production to be isotropic in space and we have obtained it by integrating on all the possible directions. Instead, actually, the strong magnetic field deflects the electrons and the positrons in such a way their ``macroscopic'' motion is along the magnetic field itself (of course, the particles also rotate in the plane perpendicular to the magnetic field but along circles with a microscopic radius). Another reason of the overestimate is the fact that we have not considered here the backreaction of the created pairs on the existing magnetic field. This backreaction reduces the magnetic field strength (and then the final pair production yield) because the created electrons and positrons rotate in such a way as to produce a magnetic field in the opposite direction of the strong magnetic field $B(t)$. It must also be said that the value $\mathcal{E}\sim 10^{57}\;\text{erg}$ would be too large with respect to the typical energy carried by a gamma-ray burst which is $\sim 10^{51}-10^{54}\;\text{erg}$. In fact, as it has been shown in \cite{Aksenov}, at so high luminosities as those considered here $\sim\mathcal{E}/\tau\sim 10^{61}\;\text{erg/s}$, almost all the pairs produced annihilate into photons. Concerning this fact, we have studied the spectrum of the photons produced in Minkowski spacetime as a consequence of the annihilation of the pairs created \cite{DiPiazza3} or as synchrotron radiation \cite{DiPiazza5}. Analogously to what we have seen in \cite{DiPiazza3}, we expect that the annihilation spectrum is again peaked around the electron mass but less sharply because the creation of high-energy pairs is here less suppressed than as in Minkowski spacetime. Also, we have seen in \cite{DiPiazza5} that the spectrum of the photons emitted as synchrotron radiation by the pairs already created has some qualitative similarities to those of gamma-ray bursts but that it decreases at higher energies faster than $\omega^{-3}$ which is the typical behaviour of the high-energy part of the gamma-ray bursts spectra \cite{GRB}. In the present case we expect qualitatively a less fast decreasing because the high-energy pair production probability scales as $E^{-3}$ instead that as $E^{-4}$ as in Minkowski spacetime.
%
%
\section{Conclusions and final considerations}
\label{V}
\setcounter{equation}{0}
\renewcommand{\theequation}{V.\arabic{equation}}
In this paper we have calculated the probability that an electron-positron pair is created in the presence of a strong magnetic field with time-varying strength and of a constant and uniform gravitational field parallel to the magnetic field and represented by the Rindler spacetime metric. We have compared this probability with the analogous one calculated in Minkowski spacetime and we have shown how large are the effects of the gravitational field. From a physical point of view this can be expected because the pair is created here in the presence of a \emph{negative} gravitational potential in such a way the rest mass of the electron does no more represent an energetic lower limit. Concerning this fact, we observe that the presence of the 
gravitational field makes possible the creation of $e^-\text{-}e^+$ pairs 
that cannot fly to infinity because they do not have enough energy. What we really expect is that the 
charged particles created with such energies annihilate inside 
the gravitational field giving rise to photons which may fly away. So, 
in the presence of the gravitational field, the 
spectrum of photons produced through annihilation extends to low frequencies. Actually, this spectrum is also expected qualitatively to be broadened at higher frequencies with respect to the Minkowski spacetime case because high-energy pairs are more easily created in the presence of the gravitational field. In fact, being $E$ the electron (positron) energy here the production probability per unit energy scales as $E^{-3}$ while in Minkowski spacetime as $E^{-4}$.

We have also pointed out a qualitative different dependence of the production probability on the magnetic field strength $B$: here it is proportional to $B^{5/2}$ while in Minkowski spacetime to $B^{3/2}$. In this respect we have observed a huge enhancement in the production probability with respect to the Minkowski spacetime case. Concerning this effect of the gravitational field, it is worth considering the ``intermediate'' situation in which a gravitational field is present in the production region but it is not so strong as in the vicinity of the event horizon of a black hole. This situation happens, for example, if the pair is imagined to be produced around a neutron star. We have already considered this case in \cite{DiPiazza4} and we refer the reader to that reference for details. We have seen there that, in the case of a neutron stars, the effects of the gravitational field can be treated perturbatively. In the different configuration in which the magnetic field was \emph{rotating} we have shown that the presence of the gravitational field in that case enhances the production of pairs but does not change the scaling of the production probability with respect to the magnetic field. This conclusion is not unexpected in a perturbative regime and it is sensible to affirm that it also holds for the present case of magnetic field varying only in strength. In this respect the conclusion is that the strong enhancement in the pair production yield we have obtained here is due to the non-perturbative effects of the gravitational field and it only occurs for the case of objects with an event horizon. Correspondingly, by means of a numerical estimate we have shown that the order of magnitude of the energy produced through this mechanism is very large and (at most) $\sim 10^{57}\;\text{erg}$ in the physical conditions previously discussed.

We want to conclude by comparing qualitatively the pair production mechanism discussed in the paper with the \emph{direct} pair production by a gravitational field. In fact, as we have said in the Introduction, many works have been done about the production of particles (photons, $e^-\text{-}e^+$ pairs) in the presence of a time-depending gravitational field \cite{Stellare,Cosmologico}\footnote{Actually, the creation of particles by a strong gravitational field is not bounded to the case of collapsing objects, but it can also happens for static situations (see, for example, the production of particles around the so-called ``eternal black holes'' in \cite{Birrell}).}. We quote, in particular, the seminal paper \cite{Hawking2} by Hawking where it was shown for the first time that a collapsing black hole with mass $M$ emits electrons (positrons) with a thermal energy spectrum proportional to the Fermi-Dirac factor
\begin{equation}
\label{Haw}
\frac{1}{\exp(\omega/k_BT)+1}
\end{equation}
where $\omega$ is the energy of the electron and $k_BT=1/(8\pi MG)=1/(4\pi r_G)$ with $k_B$ the Boltzmann constant. From the previous equation it is clear that the main emission of particles takes place for wavelengths of the order of the Schwarzschild radius of the emitting object. Now, as it has been already stated, we have studied the creation of particles with microscopic wavelengths of the order of $\lambdabar=1/m$, so the process under study takes places in a region of energies where the emission for pure gravitational effects is very small for every astrophysical object. The situation could become different when the particle production is primed by a rotating black hole. In this case, the quantization of a field in the corresponding Kerr metric is a very complicated issue because the definition itself of a vacuum is problematic \cite{Ottewill}. Nevertheless, also in \cite{Hawking2}, it is argued that the energy spectrum of the particle created is the same as in Schwarzschild case [see Eq. (\ref{Haw})] but with the particle energy $\omega$ substituted by $\omega-m_l\Omega_K$ where $m_l$ is the quantum number of the particle angular momentum component along the black hole rotational axis and $\Omega_K$ is the angular frequency of the black hole. In this way, we can conclude qualitatively that the possibility of coexistence of the two emission mechanisms could be found only in the case of really huge values of the angular momentum of the emitted particle but a quantitative estimate would involve detailed and non trivial calculations.
%
%
\appendix
\section*{Appendix}
\setcounter{equation}{0}
\renewcommand{\theequation}{A\arabic{equation}}
In this appendix we will impose that the spinor (\ref{u_d}) is continuous at 
$z=b$ and that its norm is unit. As a result, we will discretize 
the energies $E$ and determine the two factors $N^{(<)}_{n,n_d,n_g,\sigma}$ and 
$N^{(>)}_{n,n_d,n_g,\sigma}$ appearing 
in Eq. (\ref{u_d}). The continuity condition is satisfied if
\begin{equation}
\label{B_C}
N^{(<)}_{n,n_d}I_{1/2+2iE_{n,n_d}r_G}(4k_{n_d}b)=
N^{(>)}_{n,n_d}K_{1/2+2iE_{n,n_d}r_G}(4k_{n_d}b)
\end{equation}
where we pointed out that $N^{(<)}_{n,n_d}$ and 
$N^{(>)}_{n,n_d}$ can not depend on $n_g$ and $\sigma$ and that the energies 
depend on a new integer quantum number $n$.
Since $k_{n_d}b\ll 1$ we can use the approximated expressions 
of the modified Bessel functions near the origin \cite{Abramowitz}
\begin{align}
\label{I_a}
I_{1/2+2iE_{n,n_d}r_G}(4k_{n_d}b) & \sim \frac{1}{\Gamma(3/2+2iE_{n,n_d}r_G)}
(2k_{n_d}b)^{1/2+2iE_{n,n_d}r_G},\\
\label{K_a}
K_{1/2+2iE_{n,n_d}r_G}(4k_{n_d}b) & \sim \frac{\Gamma(1/2+2iE_{n,n_d}r_G)}{2}
(2k_{n_d}b)^{-1/2-2iE_{n,n_d}r_G}
\end{align}
and Eq. (\ref{B_C}) becomes
\begin{equation}
\label{B_C_2}
\frac{N^{(>)}_{n,n_d}}{N^{(<)}_{n,n_d}}
\left(\frac{1}{2}+2iE_{n,n_d}r_G\right)
\Gamma^2\left(\frac{1}{2}+2iE_{n,n_d}r_G\right)
\frac{(2k_{n_d}b)^{-1-4iEr_G}}{2}=1
\end{equation}
where the property $\Gamma (z+1)=z\Gamma (z)$ has been used. By equating 
the modulus and the phase of the left and right hand sides of Eq. (\ref{B_C_2}) we 
obtain the two real conditions
\begin{align}
\label{N_m}
& N^{(<)}_{n,n_d}=\frac{\pi}{8k_{n_d}b}\frac{\sqrt{1+(4E_{n,n_d}r_G)^2}}
{\cosh(2\pi E_{n,n_d}r_G)}N^{(>)}_{n,n_d}, \\
\label{E_n}
& \arctan\left(4E_{n,n_d}r_G\right)+2\arg{\left(\Gamma
\left(\frac{1}{2}+2iE_{n,n_d}r_G\right)\right)}-
4E_{n,n_d}r_G\log(2k_{n_d}b)=2n\pi && n=0,\pm 1,\ldots
\end{align}
where the following property of the $\Gamma$ function 
has been used \cite{Abramowitz}: 
\begin{align}
\label{mod_gamma}
\Gamma\left(\frac{1}{2}+i\xi\right)\Gamma\left(\frac{1}{2}-i\xi\right)=
\left\vert\Gamma\left(\frac{1}{2}+i\xi\right)\right\vert^2=
\frac{\pi}{\cosh(\pi\xi)}
&& \text{with $\xi\in \mathbb{R}$}.
\end{align}
The condition (\ref{E_n}) determines the allowed discrete energies while, 
in order to determine $N^{(>)}_{n,n_d}$, we have to require the following 
normalization condition [see the expression (\ref{s_p_f}) of the 
scalar product]:
\begin{equation}
64\int d\mathbf{r} u_{n,n_d,n_g,\sigma}^{\dag}
(\mathbf{r})u_{n,n_d,n_g,\sigma}(\mathbf{r})=1.
\end{equation}
It can be seen that the previous condition 
is equivalent to require that
\begin{equation}
\label{norm}
\begin{split}
64\frac{k_{n_d}r_G\cosh(2E_{n,n_d}r_G)}{4\pi^2}
&\left[N^{(<)\, 2}_{n,n_d}
\int_0^bdz\left\vert I_{1/2+2iE_{n,n_d}r_G}(4k_{n_d}z)\right\vert^2+\right.\\
&\left.+N^{(>)\, 2}_{n,n_d}
\int_b^{\infty}dz\left\vert K_{1/2+2iE_{n,n_d}r_G}(4k_{n_d}z)\right\vert^2\right]=1.
\end{split}
\end{equation}
By using the approximated expressions (\ref{I_a}) calculated in $4k_{n_d}z$ 
the first integral gives
\begin{equation}
\label{int_0_b_I}
\int_0^bdz\left\vert I_{1/2+2iE_{n,n_d}r_G}(4k_{n_d}z)\right\vert^2
\simeq\frac{1}{k_{n_d}}
\frac{\cosh(2\pi E_{n,n_d}r_G)}{1+(4E_{n,n_d}r_G)^2}
\frac{(k_{n_d}b)^2}{\pi}.
\end{equation}
The second integral can be evaluated by using the following identity
\begin{equation}
\label{int_b_inf}
\begin{split}
\int_b^{\infty}dz\left\vert K_{1/2+2iE_{n,n_d}r_G}(4k_{n_d}z)\right\vert^2=
\lim_{\epsilon\to 0}&\left[\int_0^{\infty}dz(4k_{n_d}z)^{\epsilon}
\left\vert K_{1/2+2iE_{n,n_d}r_G}(4k_{n_d}z)\right\vert^2-\right.\\
&-\left.\int_0^bdz(4k_{n_d}z)^{\epsilon}
\left\vert K_{1/2+2iE_{n,n_d}r_G}(4k_{n_d}z)\right\vert^2\right].
\end{split}
\end{equation}
The first integral in the right hand side of this equation is a particular 
case of the general formula \cite{Ryzhik} 
\begin{equation}
\label{int}
\begin{split}
\int_0^{\infty}dxx^{-\rho}K_{\lambda}(ax)K_{\mu}(bx) &=
\frac{2^{-2-\rho}a^{-\mu+\rho-1}b^{\mu}}{\Gamma(1-\rho)}
\Gamma\left(\frac{1-\rho+\lambda+\mu}{2}\right)
\Gamma\left(\frac{1-\rho-\lambda+\mu}{2}\right)\times\\
&\times\Gamma\left(\frac{1-\rho+\lambda-\mu}{2}\right)
\Gamma\left(\frac{1-\rho-\lambda-\mu}{2}\right)\times\\
&\times F\left(\frac{1-\rho+\lambda+\mu}{2},\frac{1-\rho-\lambda+\mu}{2};
1-\rho;1-\frac{b^2}{a^2}\right)
\end{split}
\end{equation}
where $\mathrm{Re}(a+b)>0$ and $\mathrm{Re}(\rho)<1-
\vert\mathrm{Re}(\lambda)\vert-\vert\mathrm{Re}(\mu)\vert$ and where $F(r,s;u;\xi)$ is 
the hypergeometric function.
Instead, in the second integral in the right hand side of Eq. (\ref{int_b_inf}) 
the approximated expression (\ref{K_a}) 
of the modified Bessel function calculated in $4k_{n_d}z$ 
can be used. Obviously, even if these integrals 
are both diverging in the limit $\epsilon\to 0$, 
their divergences must cancel each other because the 
left hand side of Eq. (\ref{int_b_inf}) is finite. In fact,
\begin{equation}
\begin{split}
&\lim_{\epsilon\to 0}\left[\int_0^{\infty}dz(4k_{n_d}z)^{\epsilon}
\left\vert K_{1/2+2iE_{n,n_d}r_G}(4k_{n_d}z)\right\vert^2-\int_0^bdz(4k_{n_d}z)^{\epsilon}
\left\vert K_{1/2+2iE_{n,n_d}r_G}(4k_{n_d}z)\right\vert^2\right]=\\
&=\frac{1}{4k_{n_d}}\lim_{\epsilon\to 0}
\left[\frac{2^{-2+\epsilon}}{\Gamma(1+\epsilon)}
\Gamma\left(\frac{2+\epsilon}{2}\right)
\left\vert\Gamma\left(\frac{1+\epsilon}{2}+2iE_{n,n_d}r_G\right)\right\vert^2
\Gamma\left(\frac{\epsilon}{2}\right)-\right.\\
&\left.\qquad\qquad\quad -\frac{1}{2\epsilon}
\left\vert\Gamma\left(\frac{1}{2}+2iE_{n,n_d}r_G\right)\right\vert^2
(4k_{n_d}b)^{\epsilon}\right]=\\
&=\frac{1}{4k_{n_d}}
\left\vert\Gamma\left(\frac{1}{2}+2iE_{n,n_d}r_G\right)\right\vert^2
\lim_{\epsilon\to 0}
\left[\frac{1}{2\epsilon}-\frac{1}{2\epsilon}
\exp{\left(\epsilon\log{(4k_{n_d}b)}\right)}\right]=\\
&=-\frac{1}{8k_{n_d}}
\left\vert\Gamma\left(\frac{1}{2}+2iE_{n,n_d}r_G\right)\right\vert^2\log{(4k_{n_d}b)}
\end{split}
\end{equation}
where we used the property 
$\Gamma(\epsilon/2)=2\Gamma(1+\epsilon/2)/\epsilon$. Finally, by exploiting 
Eq. (\ref{mod_gamma}) we have
\begin{equation}
\label{int_b_inf_K}
\int_b^{\infty}dz\left\vert K_{1/2+2iE_{n,n_d}r_G}(4k_{n_d}z)\right\vert^2=
-\frac{\pi}{8k_{n_d}\cosh(2E_{n,n_d}r_G)}\log\left(4k_{n_d}b\right)
\end{equation}
and, by substituting Eqs. (\ref{N_m}), (\ref{int_0_b_I}) and (\ref{int_b_inf_K}) in 
Eq. (\ref{norm}) we obtain the following expression 
of $N^{(>)}_{n,n_d}$
\begin{equation}
N^{(>)}_{n,n_d}=N^{(>)}_{n_d}=\sqrt{\frac{4\pi}{r_G[1-8\log(4k_{n_d}b)]}}.
\end{equation}
Since, at the end of the calculations the limit $b\to 0$ will be performed, we 
can give the expression of $N^{(>)}_{n_d}$ in this limit:
\begin{align}
N^{(>)}_{n_d}=\sqrt{-\frac{\pi}{2r_G\log\left(k_{n_d}b\right)}} 
&& k_{n_d}b\to 0.
\end{align}
In the same limit an easy expression of the density of the energy levels 
$\varrho\left(E_{n,n_d}\right)$ can be 
obtained. In fact, this quantity is defined as
\begin{equation}
\varrho\left(E_{n,n_d}\right)\equiv
\left\vert\frac{dn}{dE_{n,n_d}}\right\vert.
\end{equation}
Now, if $k_{n_d}b\to 0$ then Eq. (\ref{E_n}) becomes simply 
\begin{align}
\label{E_n_n_d}
E_{n,n_d}r_G\log \left(k_{n_d}b\right)=n\frac{\pi}{2} && k_{n_d}b\to 0
\end{align}
and the density of the energy levels does not depend on the energy itself:
\begin{align}
\label{rho}
\varrho\left(E_{n,n_d}\right)=\varrho_{n_d}=
-\frac{2r_G}{\pi}\log\left(k_{n_d}b\right) && k_{n_d}b\to 0.
\end{align}
Finally, with this definition the normalization factor $N^{(>)}_{n_d}$ 
in the limit $k_{n_d}b\to 0$ can be written as
\begin{align}
\label{N_M}
N^{(>)}_{n_d}=\frac{1}{\sqrt{\varrho_{n_d}}} && k_{n_d}b\to 0.
\end{align}
%
%
%

\clearpage
\end{document}